\documentclass[grl]{agu2001}

\ifgalley

\fi

\usepackage{graphicx}
\usepackage{latexsym}
\usepackage{lineno}
\usepackage{rotating}

\authorrunninghead{Mutel et al.}
\titlerunninghead{Angular Beaming of AKR}

\authoraddr{R. L. Mutel, Dept. Physics and Astronomy, University of Iowa,
Iowa City IA 52242 (robert-mutel@uiowa.edu)}
\authoraddr{I. W. Christopher, Dept. Physics and Astronomy, University of Iowa,
Iowa City IA 52242 (ivar-christopher@uiowa.edu)}

\lefthead{Mutel et al.}
\righthead{Angular beaming of AKR}

\newcommand{\fig}[1]{Fig.\hspace{0.8mm}\ref{#1}}

\begin{document}
\def\etal{et al. }

\title{Cluster Multi-spacecraft Determination of AKR Angular Beaming}

\authors{
R. L. Mutel \altaffilmark{1}
I. W. Christopher  \altaffilmark{1}
J .S. Pickett \altaffilmark{1}
}

\altaffiltext{1}
{Dept. Physics and Astronomy, University of Iowa, Iowa City IA 52242, USA.}
\begin{abstract}
{\bf Abstract.}  
Simultaneous observations of AKR emission using the four-spacecraft Cluster array were used to make the first direct measurements of the angular beaming patterns of individual bursts. By comparing the spacecraft locations and AKR burst locations, the angular beaming pattern was found to be narrowly confined to a plane containing the magnetic field vector at the source and tangent to a circle of constant latitude. Most rays paths are confined within 15\deg\ of this tangent plane, consistent with numerical simulations of AKR k-vector orientation at maximum growth rate. The emission is also strongly directed upward in the tangent plane, which we interpret as refraction of the rays as they leave the auroral cavity. The narrow beaming pattern implies that an observer located above the polar cap can detect AKR emission only from a small fraction of the auroral oval at a given location. This has important consequences for interpreting AKR visibility at a given location. It also helps re-interpret previously published Cluster VLBI studies of AKR source locations, which are now seen to be only a subset of all possible source locations. These observations are inconsistent with either filled or hollow cone beaming models. 
\end{abstract}

\begin{article}
\section{Introduction}

Determining the angular beaming characteristics of terrestrial auroral kilometric radiation (AKR) has been an important focus of AKR  investigations since its discovery more than forty years ago. \citet{gurnett:1974}  made the first analysis of AKR beaming using one year of Imp-6 and Imp-8 satellite observations of AKR emission. He found that the statistical power pattern formed a `distinct cone-shaped boundary' at a large angle to the magnetic field direction. Subsequent statistical analyses of AKR radiation patterns were fitted to either  frequency-dependent filled cone \citep{green:1977, green:1985} or hollow cone \citep{calvert:1981b,calvert:1987} beaming models, the latter suggested by studies of Jovian decametric emission \cite[e.g.,][]{dulk:1970}, whose properties are consistent with a very thin ($\sim1$\deg\ ) hollow cone.  More recently, \citet{kasaba:1997} analyzed 38 months of AKR bursts recorded on the GEOTAIL spacecraft. They could not distinguish between hollow and filled cone models, but found that the AKR illumination pattern was systematically modified both by geomagnetic activity and by season. 

These studies provide statistical descriptions of the overall sky pattern illuminated by AKR bursts, but they do not address the angular beaming pattern of individual AKR bursts. AKR emission is thought to consist of a large number of 'elementary radiation sources' with narrow frequency-time structures \citep{gurnett:1981,pottelette:2001,mutel:2006a}. Hence, instruments with inadequate frequency and/or time resolution (e.g. swept-frequency receivers) receive the sum of contributions from many spatially separated elementary sources whose frequency-time structures lie within the time-frequency resolution window of the instrument.  Also,  since most beaming studies have utilized observations from single satellites, the observed power pattern on the sky is a statistical ensemble of beams from many individual radiating sources rather than the angular beaming pattern of an isolated source. 

The wideband (WBD) instrument on the four Cluster spacecraft provides a unique opportunity to determine the angular power pattern of individual AKR bursts for the first time.  The WBD system records the received waveform directly, so that it has adequate time and frequency resolution to isolate emission from individual elementary AKR sources. In addition, since the Cluster constellation forms a 2-dimensional array on the sky,  it can simultaneously sample the burst over a range of solid angles. Finally, by measuring the differential delays between pairs of spacecraft, the locations of individual AKR bursts can also be determined \citep{mutel:2003}. This provides a spatial filter which isolates radiation from a single region and allow the array to sample individual burst power patterns. 

\section{AKR Beaming: Model predictions}

The electron cyclotron maser instability (CMI) is widely believed to be the mechanism responsible for auroral kilometric radio emission \citep[e.g.,][]{treumann:2006}. AKR sources are found in thin, low density cavities in the upward current region above the auroral zones \citep{calvert:1981a,ergun:1998} . The cavities are generally oriented tangent to the auroral oval and aligned with the magnetic field. They have small latitudinal widths (10 km -100 km) compared with their longitudinal extent, which can extend several thousand km. 

The radiation is generated most efficiently in the internal extraordinary (X) mode close to the electron gyro-frequency. The growth rate is strongly peaked for propagation nearly perpendicular to the magnetic field, as seen in both in situ observations \citep{roux:1993,ergun:1998} and in model calculations \citep{pritchett:2002,mutel:2007} . The internal X-mode cannot propagate directly,  but can be converted to external right circularly polarized X-mode radiation after upward propagation along the tangent plane, as first noted by \citet{louarn:1996a}. 
\citet{pritchett:2002}, using a 2-dimensional simulation of AKR emission in a thin cavity, also found that the emission is highly beamed in the 'along track' or longitudinal direction.  This prediction is strengthened by FAST in situ observations  in the upward current region which found that the tangential power is 100 times stronger than the perpendicular spectrum \citep{pritchett:2002}.
\section{Geometry of beamed AKR emission}

We used the wideband plasma wave instrument (WBD) \citep{gurnett:2001} on the four Cluster spacecraft in VLBI mode to determine the locations of individual AKR bursts for more than 12,000 individual AKR bursts over 39 epochs. The VLBI technique solves for AKR burst locations using measurements of differential delays between all six pairs of spacecraft \citep{mutel:2003}. The position uncertainty is typically 500-1000 km in the plane perpendicular to the source-observer line of sight.

\begin{figure}
  \begin{center}
    \includegraphics[width=3in,angle=0]{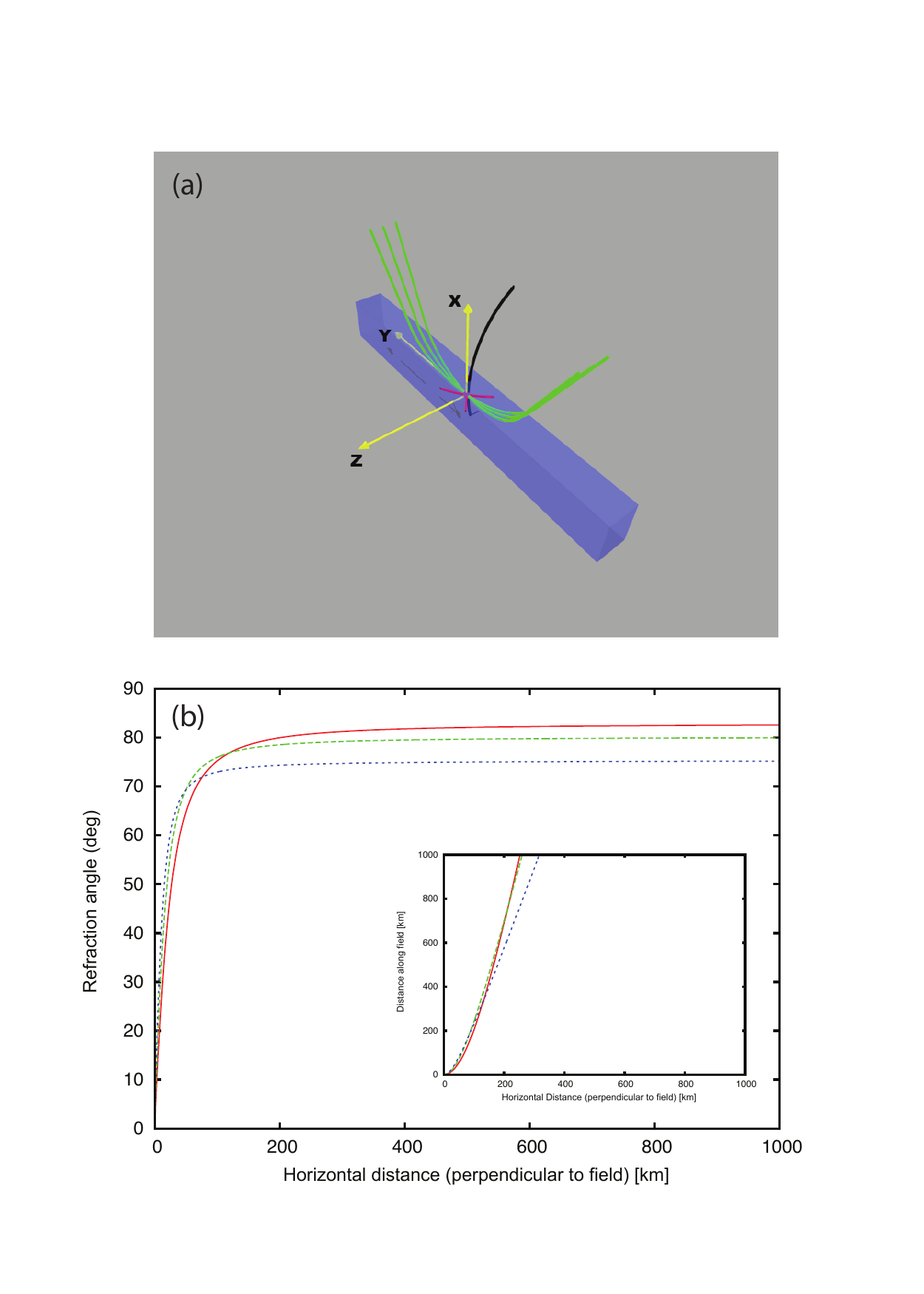}
  \end{center}
\caption{(a) Geometry of the tangent plane beaming model. The tangent plane cavity, shown in blue, is tangent to a circle of constant latitude at the source.  The green lines are escaping ray paths, the black line is the magnetic field, and yellow arrows are the ${xyz}$ coordinates system described in the text. (b) Refraction angle versus horizontal distance for rays at 125 kHz (red), 250 kHz (green), and 500 (blue) kHz initially perpendicular to the magnetic field. Ray paths are shown in the lower right inset.}
  \label{fig-raytrace}
\end{figure}

Once a source position is determined, we calculate the tangent plane coordinates of each spacecraft as seen from the source location. This is done by constructing an orthonormal 3-dimensional coordinate system with origin at the source,  $\vec{x}$ aligned outward along the magnetic field direction, $\vec{z}$ in the meridian plane pointed toward the local magnetic pole, and $\vec{y}$ = $\vec{z}\times \vec{x}$. The tangent plane latitude is the angle between the spacecraft vector and the tangent ($xy$) plane, while the longitude is the angle between the spacecraft vector projected onto the the tangent plane and $\vec{x}$. The coordinate geometry is illustrated in \fig{fig-raytrace}a.
\subsection{Confinement to tangent plane}
Once a source location is determined, we calculate the tangent plane coordinates of the Cluster spacecraft at the time of reception of the individual burst. 
Histograms of the spacecraft tangent plane latitudes and longitudes for 12,000 individual AKR bursts are shown in \fig{fig-histogram}.  The latitude distribution is strongly peaked near 0\deg\ , with more than 98\%  of solutions within 20\deg\ of the origin, and appears to be frequency-independent. The longitude distribution has been folded about the origin, since the tangent plane model is symmetric to reflection about the $\vec{x}$ axis. The histogram for all three frequencies is peaked near the origin, indicating that the ray paths are strongly refracted. 

\protect{
\begin{figure}
  \vspace{0mm}
  \begin{center}
    \includegraphics[width=3in,angle=0]{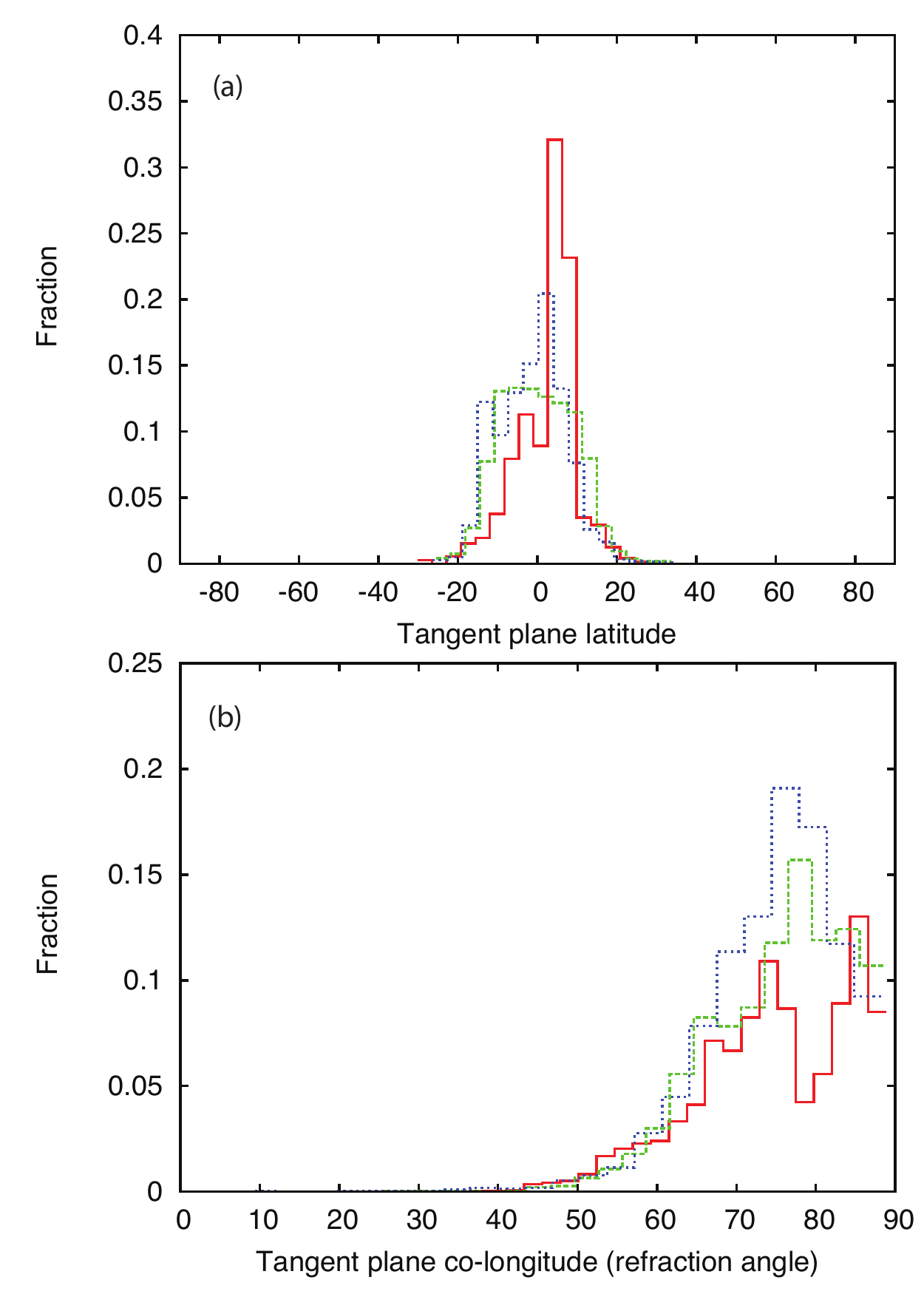}
  \end{center}
  \caption{$(a)$ Cluster spacecraft SC4 tangent plane latitude distribution at 125 kHz (red), 250 kHz (green), and 500 kHz (blue) determined from 12,000 AKR burst locations. $(b)$ Same as $(a)$, but tangent plane co-longitude distribution. }
 \label{fig-histogram}
\end{figure}
}

\subsection{Refractive effects}

As radiation leaves the auroral cavity, it is subject to refraction by the denser magnetospheric plasma. Detailed calculations of AKR propagation in the Earth's magnetosphere \citep{calvert:1987, gaelzer:1994,green:1988, schreiber:2002,burinskaya:2007,xiao:2007} predict upward refraction and/or reflection from the plasmapause. Hence, the 'tangent plane' model suggests that AKR emission is confined to a plane containing the magnetic field vector at the source, and is refracted upward (\fig{fig-raytrace}b).

In this paper we estimate the expected refraction outside the auroral cavity with a simple ray tracing calculation using the cold plasma dispersion equation. Rays at 125, 250, and 500 kHz are launched perpendicular to the magnetic field at magnetic latitude 70\deg\ from heights corresponding to the local electron gyro-frequency. We assume a dipole magnetic field, and an average electron density profile 
\begin{equation}
n_{e}(r) = 50\ {\rm{cm}}^{-3}
{
\left(
\frac{r}{2r_{e}}
\right)
}^{-2}
\end{equation}
where $r_{e}$ is the radius of the Earth. Although the electron density varies significantly with both magnetic latitude and geomagnetic activity, this profile represents an approximate long-term average \citep[cf.][Fig. 3a]{laakso:2002a}. The resulting refraction angles versus coordinate position, along with the calculated ray paths, are shown in \fig{fig-raytrace}b.
The ray tracing solution shows that the expected refraction is very strong, with average refraction angle between 70\deg\ and 80\deg\ , consistent with observed ray paths.  More realistic magnetospheric refraction models, including possible reflection from the plasmasphere, result in similar total refraction angles (e.g. \cite{xiao:2007}).

\subsection{Visibility maps}
AKR emission that is beamed in a tangent plane and refracted upward will illuminate two symmetrically displaced  azimuthal slices on a great circle belt that is the extension of the tangent plane on the sky. The belt will have a characteristic width given by the effective opening angle of the tangent plane and a longitudinal extent given by the (upward) refracted ray paths determined by the ambient magnetospheric plasma outside the cavity. For a given spacecraft location, only a fraction of all source locations on the polar cap are able to illuminate the spacecraft. Hence, for a given tangent plane opening angle and refraction angle range, we can calculate the locus of all hypothetical AKR source locations on the polar cap which can illuminate the spacecraft. We denote this locus of points as the {\it visibility map} for that spacecraft location. 


\begin{figure}[p]
\begin{sideways}
    \begin{center}
    \includegraphics[width=9in, angle=00]{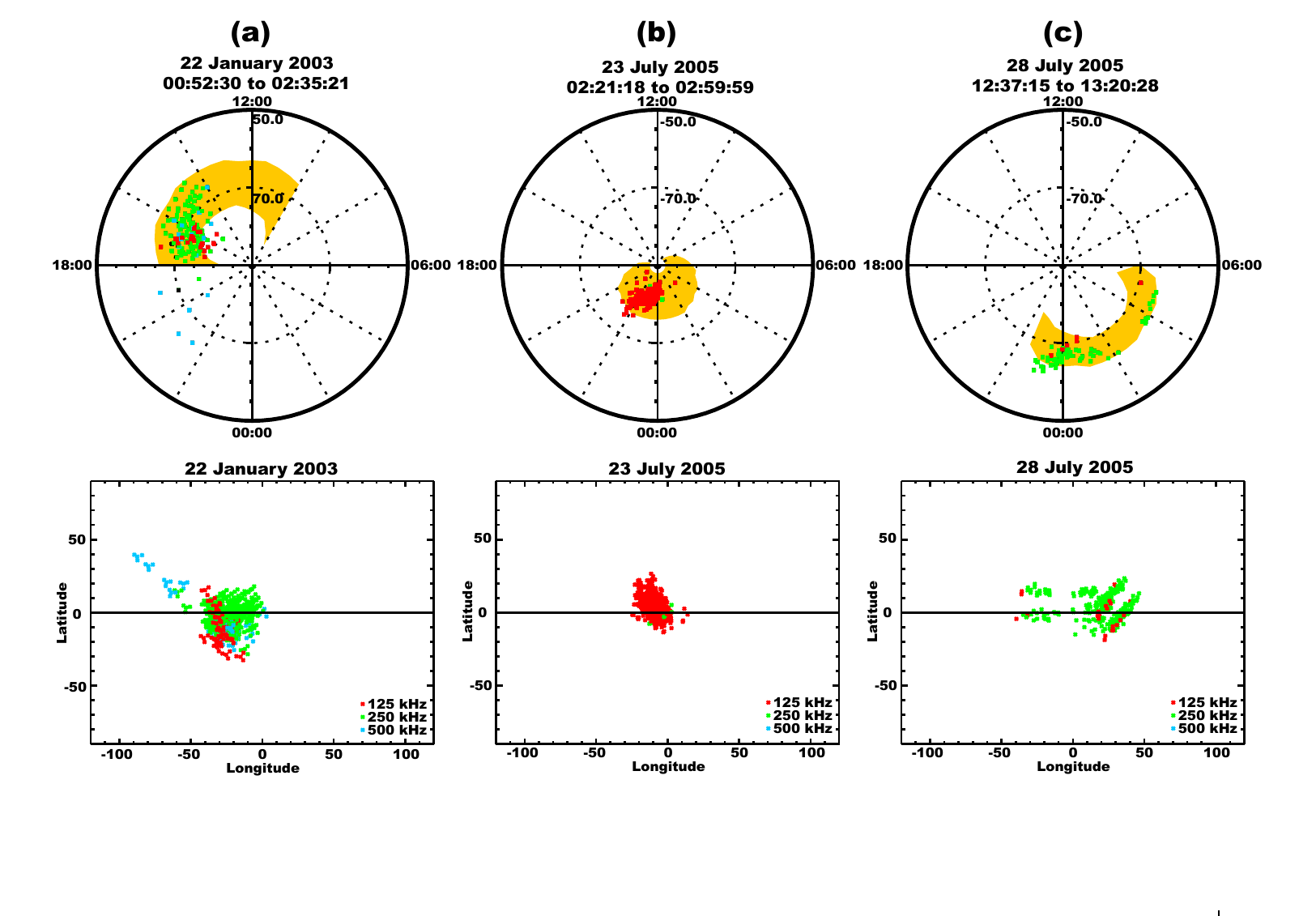}
  \end{center}
 \caption{$Upper\ panel:$ Locations of individual AKR bursts at 125 kHz (red), 250 kHz (green) and 500 kHz (blue) at epochs (a) 22 January 2003, (b) 23 Jul 2005, and (c) 28 July 2005. Mutual visibility maps for the Cluster array of spacecraft at each epoch are shown in yellow.  $Lower\ panel:$ Cluster spacecraft tangent plane coordinates as determined from each AKR burst location.}
 \label{3x2}
 \end{sideways}
\end{figure}

The upper panel of \fig{3x2} shows VLBI maps of AKR source positions for three epochs (22 Jan 2003 at 00:52 -02:35 UT, 23 July 2005 at 02:21-03:00 UT, and 28 July 2005 at 12:37-13:20 UT) mapped onto a CGM coordinate grid. Joint visibility maps of the Cluster array (i.e., the visibility area common to all four spacecraft) are shown in yellow. We computed the visibility maps assuming a tangent plane beaming pattern with a -15\deg\ to 15\deg\ latitudinal range and 0\deg\ to 45\deg\ longitude range. Note that for each epoch the AKR burst locations lie almost entirely within the visibility map. The lower panel shows the tangent plane coordinates of each spacecraft as viewed from each AKR source location. This figure provides clear confirmation that the spacecraft are located close to the tangent plane equator of each source as expected from the tangent plane beaming model. We can also see that the longitudinal positions are close to the source zenith, i.e., the ray paths suffer substantial refraction. 
\subsection {Comparison with hollow cone models}

\begin{figure}
  \begin{center}
    \includegraphics[width=3in,angle=0]{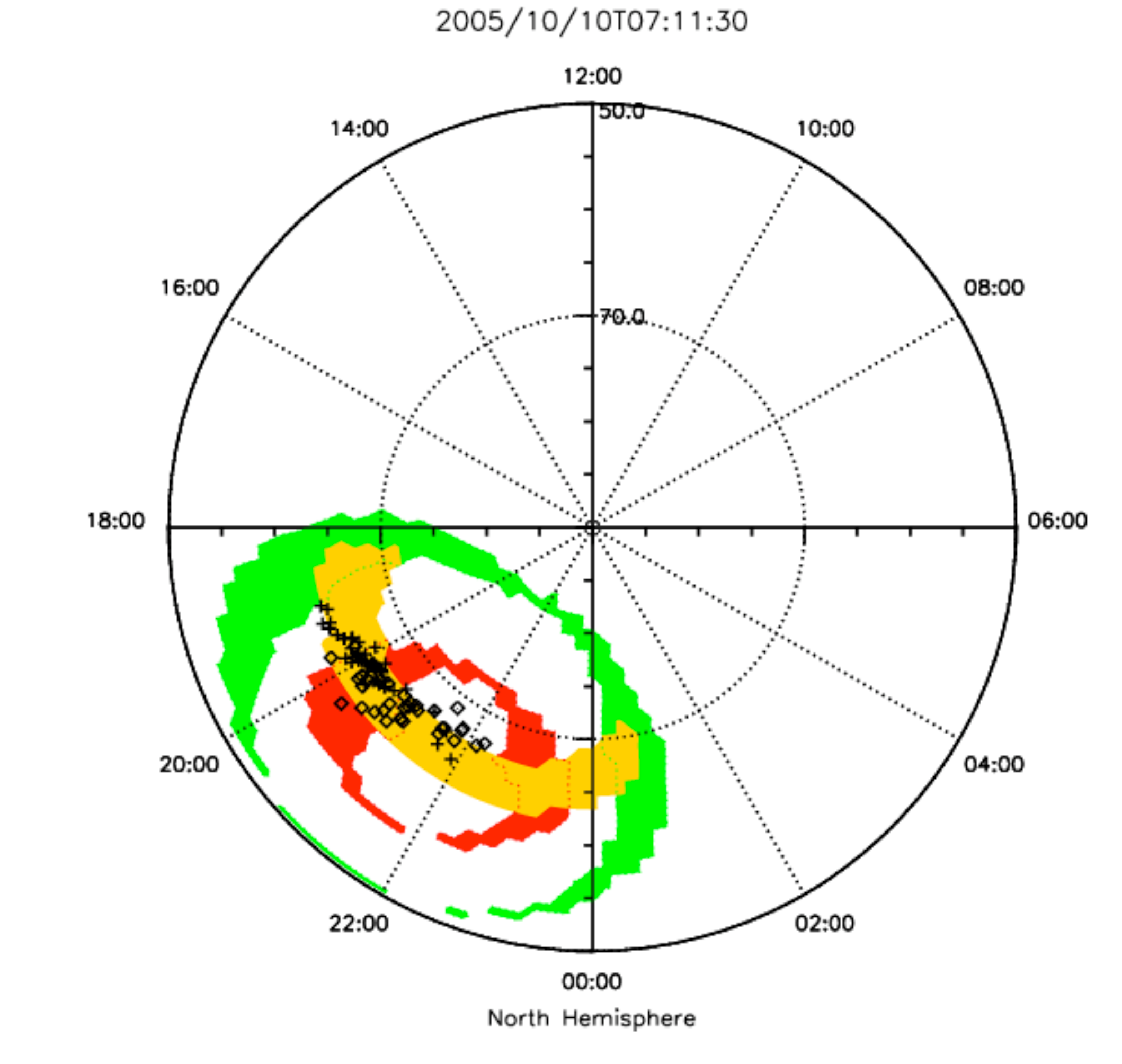}
  \end{center}
  \caption{Comparison of visibility maps for tangent plane beaming model (0\deg\ - 45\deg\ longitude, 0\deg$\pm15$\deg\ latitude, yellow) with 20\deg\ wide hollow cone models [opening angles  20\deg\ (red)  and 40\deg\ (green)]  for the 4-spacecraft Cluster array on Nov 10, 2005 at 11:30 UT. The locations of AKR burst source determined from differential delay solutions are shown as black x's (125~kHz), and squares (250~kHz).}
 \label{tp-vs-cone}
\end{figure}

Could the observations discussed in this paper also be compatible with hollow cone beaming models? In \fig{tp-vs-cone} we compare visibility maps calculated for a tangent plane beaming model (yellow) with hollow cone models having 20\deg\ opening angle (red) and 40\deg\ opening angle (green). The tangent plane model has a width of 30\deg\ while the hollow cone models have widths of 20\deg. The visibility maps were computed for Cluster spacecraft positions on Nov 10, 2005 at 11:30 UT.  The black x's (125~kHz)and squares (250~kHz) show AKR source locations determined by VLBI. It is clear that hollow cone models cannot fit the observed source locations, while the tangent plane model provides an excellent fit.

\section{Implications of the Tangent Plane Beaming Model}
The observations reported in this paper demonstrate that individual AKR bursts do not radiate in either filled or hollow cones, as previously suggested, but rather are confined to a narrow plane tangent to the source's magnetic latitude circle and containing the local magnetic field vector. The rays are also directed upward, consistent with expected refraction as rays leave the auroral cavity. This geometry confirms the numerical models of \citet{louarn:1996a} and \citet{pritchett:2002} predicting longitudinal propagation. It also implies that AKR observations from remote locations sample only a small part of the auroral oval from any given location. For example, the maps of AKR emission published by \citet{mutel:2003,mutel:2004} likely represent only a small fraction of the total extent of AKR emission on the auroral oval.

\begin{acknowledgements}
The University of Iowa acknowledges the support of NASA Goddard Space Flight Center
under Grant NNX07AI24G. We are grateful to Roman Schreiber and Jan Hanasz for several fruitful discussions.
\end{acknowledgements}
\vspace{-5mm}


\begin{thebibliography}{26}
\providecommand{\natexlab}[1]{#1}
\expandafter\ifx\csname urlstyle\endcsname\relax
  \providecommand{\doi}[1]{doi:\discretionary{}{}{}#1}\else
  \providecommand{\doi}{doi:\discretionary{}{}{}\begingroup
  \urlstyle{rm}\Url}\fi

\bibitem[{\textit{{Burinskaya} and {Rauch}}(2007)}]{burinskaya:2007}
{Burinskaya}, T.~M., and J.~L. {Rauch} (2007), {Waveguide regime of cyclotron
  maser instability in plasma regions of depressed density}, \textit{Plasma
  Physics Reports}, \textit{33}, 28--37, \doi{10.1134/S1063780X07010047}.

\bibitem[{\textit{{Calvert}}(1981{\natexlab{a}})}]{calvert:1981a}
{Calvert}, W. (1981{\natexlab{a}}), {The stimulation of auroral kilometric
  radiation by type III solar radio bursts}, \textit{\grl}, \textit{8},
  1091--1094.

\bibitem[{\textit{{Calvert}}(1981{\natexlab{b}})}]{calvert:1981b}
{Calvert}, W. (1981{\natexlab{b}}), {The AKR emission cone at low frequencies},
  \textit{\grl}, \textit{8}, 1159--1162.

\bibitem[{\textit{{Calvert}}(1987)}]{calvert:1987}
{Calvert}, W. (1987), {A Jupiter Data Analysis Program (JDAP) research grant on
  wave accessibility and attributes}, \textit{Tech. rep.}

\bibitem[{\textit{{Dulk}}(1970)}]{dulk:1970}
{Dulk}, G.~A. (1970), {Characteristics of Jupiter's Decametric Radio Source
  Measured with Arc-Second Resolution}, \textit{\apj}, \textit{159}, 671--+.

\bibitem[{\textit{{Ergun} et~al.}(1998)}]{ergun:1998}
{Ergun}, R.~E., et~al. (1998), {FAST satellite wave observations in the AKR
  source region}, \textit{\grl}, \textit{25}, 2061--2064,
  \doi{10.1029/98GL00570}.

\bibitem[{\textit{{Gaelzer} et~al.}(1994)\textit{{Gaelzer}, {Ziebell}, and
  {Schneider}}}]{gaelzer:1994}
{Gaelzer}, R., F.~L. {Ziebell}, and R.~S. {Schneider} (1994), {Ray tracing
  studies on auroral kilometric radiation in finite width auroral cavities},
  \textit{\jgr}, \textit{99}, 8905--8916.

\bibitem[{\textit{{Green}}(1988)}]{green:1988}
{Green}, J.~L. (1988), {Ray tracing planetary radio emissions}, in
  \textit{Planetary Radio Emissions II}, edited by H.~O. {Rucker}, S.~J.
  {Bauer}, and B.~M. {Pedersen}, pp. 355--379.

\bibitem[{\textit{{Green} and {Gallagher}}(1985)}]{green:1985}
{Green}, J.~L., and D.~L. {Gallagher} (1985), {The detailed intensity
  distribution of the AKR emission cone}, \textit{\jgr}, \textit{90},
  9641--9649.

\bibitem[{\textit{{Green} et~al.}(1977)\textit{{Green}, {Gurnett}, and
  {Shawhan}}}]{green:1977}
{Green}, J.~L., D.~A. {Gurnett}, and S.~D. {Shawhan} (1977), {The angular
  distribution of auroral kilometric radiation.}, \textit{\jgr}, \textit{82},
  1825--1838.

\bibitem[{\textit{Gurnett and Anderson}(1981)}]{gurnett:1981}
Gurnett, D., and R.~Anderson (1981), The kilometric radio emission spectrum:
  Relatonship to auroral acceleration processes, \textit{Geophysicsal Monogrpah
  Series}, \textit{25}, 341--350.

\bibitem[{\textit{{Gurnett}}(1974)}]{gurnett:1974}
{Gurnett}, D.~A. (1974), {The earth as a radio source: terrestrial kilometric
  radiation.}, \textit{\jgr}, \textit{79}, 4227--4238.

\bibitem[{\textit{{Gurnett} et~al.}(2001)}]{gurnett:2001}
{Gurnett}, D.~A., et~al. (2001), {First results from the Cluster wideband
  plasma wave investigation}, \textit{Annales Geophysicae}, \textit{19},
  1259--1272.

\bibitem[{\textit{{Kasaba} et~al.}(1997)\textit{{Kasaba}, {Matsumoto},
  {Hashimoto}, and {Anderson}}}]{kasaba:1997}
{Kasaba}, Y., H.~{Matsumoto}, K.~{Hashimoto}, and R.~R. {Anderson} (1997),
  {Angular distribution of auroral kilometric radiation observed by the GEOTAIL
  spacecraft}, \textit{\grl}, \textit{24}, 2483--+.

\bibitem[{\textit{{Laakso} et~al.}(2002)\textit{{Laakso}, {Pfaff}, and
  {Janhunen}}}]{laakso:2002a}
{Laakso}, H., R.~{Pfaff}, and P.~{Janhunen} (2002), {Polar observations of
  electron density distribution in the Earth's magnetosphere. 1. Statistical
  results}, \textit{Annales Geophysicae}, \textit{20}, 1711--1724.

\bibitem[{\textit{{Louarn} and {Le Qu{\'e}au}}(1996)}]{louarn:1996a}
{Louarn}, P., and D.~{Le Qu{\'e}au} (1996), {Generation of the Auroral
  Kilometric Radiation in plasma cavities-II. The cyclotron maser instability
  in small size sources}, \textit{Plan. Sp. Sci.}, \textit{44}, 211--224.

\bibitem[{\textit{{Mutel} et~al.}(2004)\textit{{Mutel}, {Gurnett}, and
  {Christopher}}}]{mutel:2004}
{Mutel}, R., D.~{Gurnett}, and I.~{Christopher} (2004), {Spatial and temporal
  properties of AKR burst emission derived from Cluster WBD VLBI studies},
  \textit{Annales Geophysicae}, \textit{22}, 2625--2632.

\bibitem[{\textit{{Mutel} et~al.}(2003)\textit{{Mutel}, {Gurnett},
  {Christopher}, {Pickett}, and {Schlax}}}]{mutel:2003}
{Mutel}, R.~L., D.~A. {Gurnett}, I.~W. {Christopher}, J.~S. {Pickett}, and
  M.~{Schlax} (2003), {Locations of auroral kilometric radiation bursts
  inferred from multispacecraft wideband Cluster VLBI observations. 1:
  Description of technique and initial results}, \textit{Journal of Geophysical
  Research (Space Physics)}, \textit{108}, 8--1, \doi{10.1029/2003JA010011}.

\bibitem[{\textit{{Mutel} et~al.}(2006)\textit{{Mutel}, {Menietti},
  {Christopher}, {Gurnett}, and {Cook}}}]{mutel:2006a}
{Mutel}, R.~L., J.~D. {Menietti}, I.~W. {Christopher}, D.~A. {Gurnett}, and
  J.~M. {Cook} (2006), {Striated auroral kilometric radiation emission: A
  remote tracer of ion solitary structures}, \textit{Journal of Geophysical
  Research (Space Physics)}, \textit{111}, 10,203--+,
  \doi{10.1029/2006JA011660}.

\bibitem[{\textit{{Mutel} et~al.}(2007)\textit{{Mutel}, {Peterson}, {Jaeger},
  and {Scudder}}}]{mutel:2007}
{Mutel}, R.~L., W.~M. {Peterson}, T.~R. {Jaeger}, and J.~D. {Scudder} (2007),
  {Dependence of cyclotron maser instability growth rates on electron velocity
  distributions and perturbation by solitary waves}, \textit{Journal of
  Geophysical Research (Space Physics)}, \textit{112}, 7211--+,
  \doi{10.1029/2007JA012442}.

\bibitem[{\textit{{Pottelette} et~al.}(2001)\textit{{Pottelette}, {Treumann},
  and {Berthomier}}}]{pottelette:2001}
{Pottelette}, R., R.~A. {Treumann}, and M.~{Berthomier} (2001), {Auroral plasma
  turbulence and the cause of auroral kilometric radiation fine structure},
  \textit{\jgr}, \textit{106}, 8465--8476, \doi{10.1029/2000JA000098}.

\bibitem[{\textit{{Pritchett} et~al.}(2002)\textit{{Pritchett}, {Strangeway},
  {Ergun}, and {Carlson}}}]{pritchett:2002}
{Pritchett}, P.~L., R.~J. {Strangeway}, R.~E. {Ergun}, and C.~W. {Carlson}
  (2002), {Generation and propagation of cyclotron maser emissions in the
  finite auroral kilometric radiation source cavity}, \textit{Journal of
  Geophysical Research (Space Physics)}, \textit{107}, 13--1,
  \doi{10.1029/2002JA009403}.

\bibitem[{\textit{{Roux} et~al.}(1993)}]{roux:1993}
{Roux}, A., et~al. (1993), {Auroral kilometric radiation sources - In situ and
  remote observations from Viking}, \textit{\jgr}, \textit{98}, 11,657--+.

\bibitem[{\textit{{Schreiber} et~al.}(2002)\textit{{Schreiber}, {Santolik},
  {Parrot}, {Lefeuvre}, {Hanasz}, {Brittnacher}, and {Parks}}}]{schreiber:2002}
{Schreiber}, R., O.~{Santolik}, M.~{Parrot}, F.~{Lefeuvre}, J.~{Hanasz},
  M.~{Brittnacher}, and G.~{Parks} (2002), {Auroral kilometric radiation source
  characteristics using ray tracing techniques}, \textit{Journal of Geophysical
  Research (Space Physics)}, \textit{107}, 20--1, \doi{10.1029/2001JA009061}.

\bibitem[{\textit{{Treumann}}(2006)}]{treumann:2006}
{Treumann}, R.~A. (2006), {The electron cyclotron maser for astrophysical
  application}, \textit{Astron. Astrophys. Rev.}, \textit{13}, 229--315,
  \doi{10.1007/s00159-006-0001-y}.

\bibitem[{\textit{{Xiao} et~al.}(2007)\textit{{Xiao}, {Chen}, {Zheng}, and
  {Wang}}}]{xiao:2007}
{Xiao}, F., L.~{Chen}, H.~{Zheng}, and S.~{Wang} (2007), {A parametric ray
  tracing study of superluminous auroral kilometric radiation wave modes},
  \textit{Journal of Geophysical Research (Space Physics)}, \textit{112},
  10,214--+, \doi{10.1029/2006JA012178}.

\end{thebibliography}

\end{article}

\end{document}